\documentclass[aps,prl,preprint]{revtex4}
\usepackage{epsfig}
\usepackage{axodraw}
\usepackage{graphicx}

\newcommand{\gsim}{\lower.7ex\hbox{$\;\stackrel{\textstyle>}{\sim}\;$}}
\newcommand{\lsim}{\lower.7ex\hbox{$\;\stackrel{\textstyle<}{\sim}\;$}}

\begin{document}

\title{
TeV leptogenesis in Z-prime models and its collider probe}

\author{Eung Jin Chun}
\affiliation{Korea Institute for Advanced Study, Seoul 130-722,
Korea, and\\
Institute for Particle Physics Phenomenology, University of
Durham, DH1 3LE, UK}

\date{\today}

\begin{abstract}
We show that  the $U(1)^\prime$ models linked with the seesaw
mechanism at TeV scale can lead to a successful baryogenesis
through soft leptogenesis with a resonant behavior in the $B$
parameter. Such a consideration constrains the $Z^\prime$ mass to
be larger than $2-3$ TeV depending on the seesaw scale and the
spharelon rate. Together with multi-TeV $Z^\prime$, large
sneutrino-antisneutrino mixing and CP violating phenomena required
by TeV leptogenesis could be searched for in future colliders by
observing  the distinct  same-sign dilepton--dichargino as well as
dislepton--diHiggs signatures.
\end{abstract}

\maketitle

The observed neutrino masses and mixing could be explained by the
celebrated seesaw mechanism \cite{minko} at TeV scale, which
requires the  neutrino Yukawa coupling of the order of the
electron Yukawa coupling. Such a seesaw mechanism may well be
linked with an extended gauge symmetry spontaneously broken at low
energy. A prototype would be  the left-right symmetric model based
on the gauge group $SU(2)_L\times SU(2)_R\times U(1)_{B-L}$
\cite{lrsm}. More economically, one can think of the gauge group
$SU(2)_L\times U(1)_Y \times U(1)^\prime$ which appears as  a
low-energy sector of the grand unification; $E_6 \to SU(5)\times
U(1)_\chi \times U(1)_\psi$ where the $U(1)^\prime$ is a  certain
linear combination of $U(1)_{\chi}$ and $U(1)_\psi$
\cite{Zreview}. In this case,  three right-handed neutrinos $N$
are usually required for anomaly cancellation \cite{nonexotic}.
Then, the TeV-scale seesaw mechanism can be directly tested by
observing clean signatures of same-sign dileptons arising from the
Majorana nature of the neutrinos \cite{keung}, together with
collider searches for the associated extra gauge boson, say,
$Z^\prime$ of $U(1)^\prime$.

Leptogenesis  is an attractive feature of the seesaw mechanism,
providing a nice way of  generating the baryon asymmetry of the
Universe \cite{fukugita}. In the usual thermal leptogenesis, the
right-handed neutrino mass is required to be larger than about
$10^8$ GeV assuming the typical hierarchical mass pattern
\cite{ibarra}. For a  TeV-scale  leptogenesis to work, one could
however have highly degenerate right-handed neutrinos realizing a
resonant enhancement \cite{resonantL}.
 In supersymmetric theories, new ways
for leptogenesis arise due to lepton number violating
(right-handed) sneutrino-antisneutrino
($\tilde{N}$-$\tilde{N}^\dagger$) mixing and  CP violation in soft
supersymmetry breaking sector \cite{softL}. This mechanism
(so-called soft leptogenesis) can work within a single generation
and requires unconventionally small $B$ parameter of the lepton
number violating soft mass.  Such a situation may be realized in a
radiative way consistently with TeV seesaw mechanism \cite{chun}.
In gauge-mediated supersymmetry breaking theories, the small
gravitino mass, $m_{3/2}=(10^{-2}-10)$ eV, can be more naturally
related to the similarly small $B$  \cite{kitano}, which is indeed
in the right range for our consideration. Recently, a number of
attempts have been made to realize successful thermal leptogenesis
at TeV scale \cite{tev,king,apos}, in anticipation of future
detection.

\vskip 1ex

In this paper, we will show that the supersymmetric seesaw
mechanism associated with an extra gauge symmetry at TeV scale is
a viable option for the generation of the cosmological baryon
asymmetry, which leads to testable predictions in colliders. As
mentioned,  the existence of a new gauge interaction certainly
makes the model highly testable as $N$ and $\tilde{N}$ can be
produced in colliders through, e.g., $e^+e^-  \to NN,
\tilde{N}\tilde{N}^\dagger$, which would not be the case only with
the usually small neutrino Yukawa couplings, $h\sim10^{-6}$. In
favor of collider tests, one would like to have a larger gauge
coupling and smaller masses for the new particles. However, such a
situation may be in contradiction with the observed  baryon
asymmetry, $Y_B\sim 10^{-10}$. Indeed, the inverse process, i.e.,
the gauge annihilation, $\tilde{N} \tilde{N}^\dagger, N N \to $
light fermions, with the standard gauge coupling strength strongly
suppresses the resulting lepton asymmetry. This puts a constraint
on the new gauge boson mass, which depends sensitively on the
spharelon dynamics below the electroweak phase transition
\cite{apos,btol}. The existence of an extra gauge boson at
mutli-TeV region will imply that the resonant behavior $B\sim
\Gamma \sim 0.1$ eV and large CP violation have to occur for a
successful leptogenesis. This is an interesting possibility for
the collider search, as the model predicts testable signatures in
the $\tilde{N}$-$\tilde{N}^\dagger$ oscillation and the associated
CP violating phenomena, which may link  cosmology with collider
physics.

\vskip 1ex

For the definite discussion of the seesaw mechanism with an extra
gauge boson, let us take the $Z^\prime$ model with $U(1)_\chi$
which appears as $ SO(10) \to SU(3)_c\times SU(2)_L\times U(1)_Y
\times U(1)_\chi $, and assume the standard gauge coupling
strength, $\alpha_\chi = \alpha' \approx 1/60$. In terms of the
$SU(5)$ notation, the SM fermions and the singlet field $N$ carry
the $U(1)_\chi$ charges as follows:
$$ {\bf 10}\, ({-1 \over\sqrt{40}}),\quad {\bf \overline{5}}\,
({3\over\sqrt{40}}), \quad {\bf 1}\, ({5\over\sqrt{40}}).$$
 Furthermore,  we will suppress the family indices of
three right-handed (s)neutrinos, as the soft leptogenesis does not
require mixing among different families,  and then  calculate the
baryon asymmetry arising from a right-handed sneutrino decay with
the typical neutrino Yukawa coupling corresponding to the
atmospheric neutrino mass,  $m_\nu =0.05 \mbox{ eV} \sim  h^2
v^2/M$. The essential features of our results can be applied also
to other classes of $Z^\prime$ models.

The superpotential of the seesaw model is
\begin{equation}
 W = h L H_2 N + {1\over2} M  N N
\end{equation}
where $L, H_2$ and $N$ denote the  lepton, Higgs and right-handed
neutrino superfields, respectively.  The soft supersymmetry
breaking terms are
\begin{equation}
 V_{soft} = \left[ A h \tilde{L} H_2 \tilde{N} + {1\over2} B M
 \tilde{N}\tilde{N} + h.c.\right] + \tilde{m}^2 \tilde{N}^\dagger \tilde{N}
\end{equation}
where all the fields are understood as the scalar components of
the superfields in Eq.~(1), and $A,B, \tilde{m}$ are dimension-one
soft parameters.  To denote the corresponding fermion components,
the notation of $L$, $\tilde{H}_2$ and $N$ will be used.  Without
loss of generality, one can take $M$ and $B$ to be real and
positive. Then, the right-handed sneutrino field is written as
$\tilde{N}= (\tilde{N}_1 + i \tilde{N}_2)/\sqrt{2}$ in which two
mass eigenstates $\tilde{N}_{1,2}$ have
\begin{equation}
 M^2_{\tilde{N}_{1,2}}= M^2 + \tilde{m}^2\pm BM \,,
\end{equation}
and thus the mass-squared difference $\Delta M^2_{\tilde{N}} =
2BM$.  For $B\ll M$, the mass difference becomes $\Delta
M_{\tilde{N}} \approx BM/M_{\tilde{N}}$ where  $M^2_{\tilde{N}}
\equiv M^2+\tilde{m}^3$.  The Yukawa couplings of
$\tilde{N}_{1,2}$ can be read off from the Lagrangian
\begin{equation}
{\cal L} = h \tilde{N} L \tilde{H}_2 + h (A \tilde{N} + M
\tilde{N}^\dagger) \tilde{L} H_2 + h.c. \,.
\end{equation}
Let us now calculate the lepton and CP asymmetry $\epsilon_X$
arising from a particle $X$ decaying to the final state $F$:
$$
\epsilon_X \equiv  \frac{\Gamma(X \to F) - \Gamma(X^\dagger \to
F^\dagger) } { \Gamma(X \to F) + \Gamma(X^\dagger \to F^\dagger) }
$$
where $F = L H_2$ or $\tilde{L} \tilde{H}_2$ in our case. Using
the effective field theory approach of resumed propagators for
unstable particles  ($X=\tilde{N}_{1,2}$) \cite{resonantL}, we
find
\begin{equation}\label{lasymmetry}
\epsilon_{1,2} ={2BM M_{\tilde{N}}(\Gamma_1+\Gamma_2)\,\xi_{1,2}
\over 4B^2M^2 + M^2_{\tilde{N}} (\Gamma_{2,1}^{2} -
\Gamma_1\Gamma_2 \xi_{1,2}^2 )} { M^2_{\tilde{N}} + |A|^2 - M^2
\over M^2_{\tilde{N}} + |A|^2 + M^2 },
\end{equation}
where
\begin{eqnarray}
\xi_{1,2} &=& -{|h|^2\over 4\pi} {\mbox{Im}(A) M \over
\Gamma_{1,2} M_{\tilde{N}} } , \nonumber\\
 \mbox{and} \quad
\Gamma_{1,2} &=& {|h|^2\over 8\pi} (1+{ |A\pm M|^2\over
M^2_{\tilde{N}} }) M_{\tilde{N}} \,. \nonumber
\end{eqnarray}
Note that, in Eq.~(\ref{lasymmetry}), we have neglected the
thermal effect breaking supersymmetry \cite{softL} which is
suppressed compared to soft breaking effect by $A$ and $\tilde{m}$
when they are of the same order of  the right-handed neutrino mass
$M$.  As one can see, $\epsilon_{1,2}$ vanishes in the limit of
$A\to 0$ and $M_{\tilde{N}}\to M$  without including the
difference between the bosonic and fermionic  thermal
distributions. Note that one can have $\epsilon \sim 1$ for $B\sim
\Gamma$ and the order one phase of $A$ with $|A| \sim M,
M_{\tilde{N}}$. Taking the approximation of $\Gamma \approx h^2
M/4\pi \approx m_\nu M^2/4\pi v^2$, one obtains the resonance
condition:
\begin{equation}
\label{Bis} B\sim \Gamma\sim 0.1 {\rm eV} \left( m_\nu \over 0.05
\mbox{ eV} \right)
 \left( M\over {\rm TeV} \right)^2 .
\end{equation}

The Boltzmann equations governing the number densities of the
sneutrino fields, $Y_X$, and lepton-antilepton asymmetry, $Y_l$,
in unit of the entropy density are
\begin{eqnarray} \label{boltz}
 {d Y_X \over d z} &=& - z K \left[ \gamma_D (Y_X-Y_X^{eq}) +
 \gamma_A {(Y_X^2-Y_X^{eq\,2})\over Y_X^{eq}}  \right] \nonumber\\
 {d Y_l \over d z} &=& 2 z K \gamma_D\left[ \epsilon (Y_X-Y_X^{eq})
 - {Y_X^{eq} \over 2 Y_l^{eq} } Y_l \right]
\end{eqnarray}
where $K\equiv \Gamma/H_1$ and $H_1=1.66 \sqrt{g_*}
M^2_{\tilde{N}}/m_{Pl}$ is the Hubble parameter at the temperature
$T=M_{\tilde{N}}$.   We take the Standard Model value of
$g_*=106.75$.  In Eq.~(\ref{boltz}), $\gamma_D = K_1(z)/K_2(z)$ is
the usual contribution from the decay due to the neutrino Yukawa
coupling $h$, and $\gamma_A$ accounts for the annihilation effect
of the snuetrinos to the light Standard Model fermions mediated by
the heavy $Z^\prime$ field: $\tilde{N} \tilde{N}^\dagger \to
Z^\prime \to f\bar{f}$. The gauge annihilation contribution is
given by
\begin{equation}
\gamma_A = {5\over \pi} {\alpha_\chi^2 M_{\tilde{N}} \over K H_1}
\int^\infty_1\!\! d t\, {K_1(2zt)\over K_2(z)} {t^3 (t^2-1)^{3/2}
\over (t^2 - {1\over 4}r^2)^2 + {1\over16} u^2}
\end{equation}
where $r\equiv M_{Z^\prime}/M_{\tilde{N}}$ and $u\equiv r
\Gamma_{Z^\prime}/M_{\tilde{N}}$.  For our numerical solution, we
simplify the Boltzmann equation by taking $\Gamma=\Gamma_{1,2}$
and $\epsilon=\epsilon_{1,2}$.    Taking the estimation of
$\Gamma$ in Eq.\ (\ref{Bis}), we get $ K \sim {m_\nu/
5\!\times\!10^{-4} \mbox{eV} } \sim 100 $ for $m_\nu = 0.05$ eV,
independently of $M_{\tilde{N}}$. Note that the annihilation
effect gets very strong with the factor of $M_{\tilde{N}}/H_1 \sim
10^{15}$ for the TeV-scale mass $M_{\tilde{N}}$. As one can expect
from such large $K$ and even larger gauge annihilation effect, the
right-handed sneutrinos follow closely the thermal equilibrium
distribution until very low temperature, and the lepton asymmetry
freezes out at large $z$, that is, $T <100$ GeV.  Such behavior is
shown in Fig.\ 1 where  the evolution of the lepton asymmetry,
$\log(Y_l/\epsilon)$, is plotted.  One finds that $Y_l$ becomes
larger for a larger $M_{Z^\prime}$ which suppresses more the
annihilation effect, and finally approaches an asymptotic value
for which the annihilation effect eventually drops out due to the
Boltzmann suppression.

\begin{figure}
\includegraphics[width=0.65\textwidth]{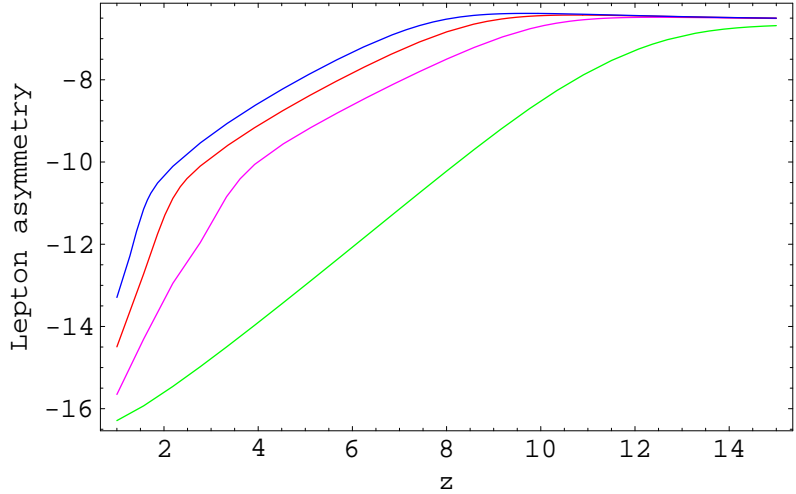}
\caption{\label{fig1} The lepton asymmetry in unit of $\epsilon$
is calculated as a function of  $z= M_{\tilde{N}}/T$ with
$M_{\tilde{N}}=0.3$ TeV and $K=500$. The curves are for
$M_{Z^\prime}=1,2,3$ and $4$ TeV from below.  }
\end{figure}

Let us note that the lepton asymmetry is still increasing during
the electroweak phase transition $T\sim 100$ GeV which is well
before the asymptotic value is reached. Therefore, it is important
to include the spharelon effect after the phase transition to
obtain the right amount of  baryon asymmetry.  To do this, let us
define $\tilde{Y}_l= Y_L - Y_B$ where the $\tilde{Y}_l$ is the
solution of the above leptogenesis equations (\ref{boltz}) as
given in Fig.\ 1. Then, it can be convoluted to the baryon
asymmetry as
\begin{equation}
 Y_B(z) = -\int_0^z\! dy\, A_1(y) \tilde{Y}_l(y) \exp[- \int^z_y\!
 dx\,
 A_2(x)]
\end{equation}
where
\begin{eqnarray}
A_1(z)  &=& z {\Gamma_{sp} \over H_1} \left( {28\over 51}
 + {108\over 561} {v^2(T)\over T^2}  \right) \nonumber\\
A_2(z) &=& z {\Gamma_{sp} \over H_1} \left( {79\over 51} +
{333\over 561} {v^2(T)\over T^2}\right) .\nonumber
\end{eqnarray}
Here $\Gamma_{sp}$ is spharelon interaction rate and $v(T)=v_0
(1-T^2/T_c^2)^{1/2}$ with $v_0=246$ GeV.  Above the electroweak
phase transition $T>T_c$, $v(T)=0$ and $\Gamma_{sp}$ can be taken
to be infinitely large so that the standard result, $Y_B =
-{28\over 51} Y_L = -{28\over 79} \tilde{Y}_l$, is recovered.
Below $T_c$, several calculations for $\Gamma_{sp}$ have been made
within the validity range of $ M_W(T) \ll T \ll M_W(T)/\alpha_w$
\cite{btol}, which indicates that the spharelon interaction is
still very active just below $T_c$ and its freeze-out happens
somewhat later \cite{apos}.  One finds that $\Gamma_{sp}$ is an
extremely steep function of $T$ and thus it is a fairly good
approximation to calculate the final baryon asymmetry as follows:
\begin{equation} \label{basymmetry}
Y_B \approx -{A_1\over A_2} \tilde{Y}_l\Big|_{T_{sp}} \approx
-{1\over3} \tilde{Y}_l \Big|_{T_{sp}}
\end{equation}
where $T_{sp}$ is the spharelon freeze-out temperature.  There is
an uncertainty in determining $T_{sp}$ which comes from the
limited knowledge in calculating the spharelon rate. The most
uncertain quantity, called $\kappa$, is known to  lie in the
range; $\kappa=(10^{-4}-1)$ \cite{btol}.  Adopting the parameters
in Ref.~\cite{apos}, we find $T_{sp}=80-90$ GeV which is in the
region of   $T>M_W(T)$ where the above spharelon calculation may
well be extended.

Following the above prescription, the baryon asymmetry,
$\log(Y_B/\epsilon)$, is calculated in Fig.\ 2 as a function of
$M_{Z^\prime}$ for various values of $M_{\tilde{N}}$, taking
$T_{sp}=90$ GeV. As one can see, $Y_B$ depends strongly on the
masses, $M_{Z^\prime}$ and $M_{\tilde{N}}$, while we find it
almost unaffected by the variation of $K$ in the range:
$K=10-10^3$. Requiring $Y_B/\epsilon>10^{-10}$, one obtains the
lower bound: $M_{Z^\prime} > (2.3-3)$ TeV for $M_{\tilde{N}}=
(0.3-0.9)$ TeV. Taking $T_{sp}=80$ GeV, the bound becomes weaker:
$M_{Z^\prime} > (2.1-2.6)$ TeV.

\begin{figure}
\includegraphics[width=0.65\textwidth]{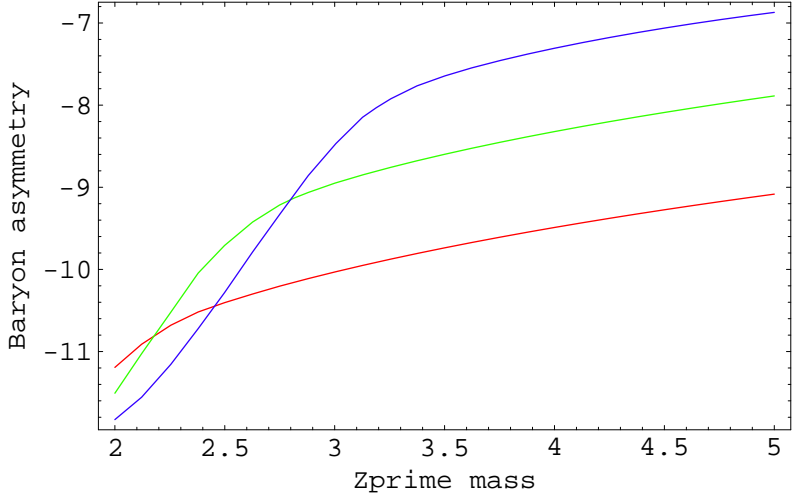}
\caption{\label{fig2} The baryon asymmetry in unit of $\epsilon$
is shown with respect to $M_{Z^\prime}/\mbox{TeV}$ for $K=100$.
The curves are for $M_{\tilde{N}}=0.3, 0.6$ and $0.9$ TeV from
below on the right-hand side.}
\end{figure}

\vskip 1ex

Such a multi-TeV $Z^\prime$ is  beyond the  Tevatron II limit, but
could be within the reach of LHC or future LC. Future collider
sensitivities for $Z^\prime$ and also for $N$ ($\tilde{N}$) have
been analyzed in the literature. For LHC with the accumulated
luminosity 100 fb$^{-1}$, a few hundred to a few $Z^\prime \to NN$
events can be observed  for $M_{Z^\prime} = 3-5$ TeV. For LC at
$E_{cm} = 3$ TeV with the luminosity 1000 fb$^{-1}$,  $10^5$
events will be obtained for $M_{Z^\prime}=3$ TeV \cite{futureZ}.
With such large number of events, we will be able to study the
properties of $N$ and $\tilde{N}$. To cover more parameter space
of our cosmological interest, an up-graded version of LHC or a
higher energy option for LC will certainly be useful.

Apart from the well-studied  $Z^\prime$ signals and  dilepton
(plus diHiggs) final states  from $Z^\prime \to NN$ \cite{keung},
sneutrino-antisneutrino mixing and CP violating phenomena can be
looked for to test the soft leptogenesis mechanism.   Near the
resonance point, $B\sim \Gamma$, order-one
$\tilde{N}$-$\tilde{N}^\dagger$ oscillation  effects are expected
to occur in colliders, favoring the parameter space of
$M_{Z^\prime} \lsim 3$ TeV and $\epsilon\sim 1$ which also
requires a large CP phase, $\mbox{Im}(A)$.  Following the decay
$Z^\prime \to \tilde{N} \tilde{N}^\dagger$, the final states with
a same-sign lepton (slepton) pair and chargino (charged Higgs) can
occur through oscillations:
\begin{eqnarray}
 \tilde{N} \tilde{N}^\dagger
 &\Rightarrow& \tilde{N} \tilde{N},\; \tilde{N}^\dagger
 \tilde{N}^\dagger \nonumber\\
  &\rightarrow& l^\pm l^\pm \chi^\mp\chi^\mp \\
 &\to& \tilde{l}^\pm \tilde{l}^\pm H^\mp H^\mp \,.\nonumber
\end{eqnarray}
As in the $B$-$\bar{B}$ mixing \cite{schneider}, let us define the
oscillation parameters for sneutrinos: $x\equiv \Delta
M_{\tilde{N}}/\Gamma= |B|M/\Gamma M_{\tilde{N}}$ and $y\equiv
\Delta \Gamma/2\Gamma$ where $\Gamma=(\Gamma_1+\Gamma_2)/2$ and
$\Delta \Gamma = \Gamma_1-\Gamma_2$, which can be measures in the
mixing processes as we discuss. Various observables involving the
oscillation parameters and the CP phase $\mbox{Im}(A)$ are as
follows.

(i) Oscillation from the same-sign dileptons:
$$
 {N(l^-l^-) + N(l^+l^+) \over N(l^+l^-) + N(l^-l^+)} = {2 r
 \over 1 + r^2 }
$$
where $r\equiv (x^2+y^2)/(2+x^2-y^2)$.

 (ii) Associated CP asymmetry:
$$
 {N(l^-l^-) - N(l^+l^+) \over N(l^-l^-) + N(l^+l^+)} =
 \mbox{Im}\left( h^2 A\over 4\pi B\right) .
$$

(iii) Oscillation from the same-sign disleptons:
$$
{N(\tilde{l}^-\tilde{l}^-) +
N(\tilde{l}^+\tilde{l}^+) \over N(\tilde{l}^-\tilde{l}^+) +
N(\tilde{l}^+\tilde{l}^-) } = { (1+r^2)a+ r (1+a^2) \over
(1+r^2)(1+a^2) + 2ra }
$$
where $a\equiv |A/M|^2$.

(iv) The ratio of CP asymmetries:
$$
{N(\tilde{l}^-\tilde{l}^-) - N(\tilde{l}^+\tilde{l}^+) \over
N(l^-l^-) - N(l^+l^+)} = { |A|^4-|M|^4 \over M^4_{\tilde{N}}} \,.
$$
If one measures the above quantities together with the decay rates
and the sneutrino mass, one can basically reconstruct the amount
of the cosmological lepton asymmetry given in
Eq.~(\ref{lasymmetry}) and  obtain the final baryon asymmetry from
Eq.~(\ref{basymmetry}), which could provide a way to study
cosmology in colliders.

In conclusion, it is shown that, in the low energy $U(1)^\prime$
models endowed with the seesaw mechanism,   a successful
leptogenesis can arise from supersymmetry breaking sector
realizing the resonance behavior, $B\sim \Gamma$. Such a
cosmological consideration puts a lower bound on the $Z^\prime$
mass typically at multi TeV,  assuming the standard gauge coupling
strength. The bound is sensitive to the right-handed (s)neutrino
mass as well as the spharelon rate which has a certain theoretical
uncertainty. Further understanding on the spharelon dynamics would
be required in the future confronting  collider experiments. It is
also pointed out that the soft leptogenesis mechanism of the model
can be readily tested by the future observation of not only the
$Z^\prime$ at multi-TeV range but also sneutrino-antisneutrino
mixing and CP violation leading to distinct signatures of the
same-sign dilepton (dislepton) and dichargino (diHiggs)
production.

\vskip 1ex

{\bf Acknowledgement}: The author thanks Stefano Scopel for
discussions and confirmation of some results.  He is also grateful
to Apostolos Pilaftsis and Thomas Underwood for useful comments
and criticisms.


\begin{thebibliography}{99}

\def\plb#1#2#3{Phys.\ Lett.\       {\bf B#1},  #2 (#3)}
\def\npb#1#2#3{Nucl.\ Phys.\       {\bf B#1},  #2 (#3)}
\def\prd#1#2#3{Phys.\ Rev.\        {\bf D#1},  #2 (#3)}
\def\prl#1#2#3{Phys.\ Rev.\ Lett.\ {\bf #1},   #2 (#3)}
\def\rmp#1#2#3{Rev.\ Mod.\ Phys.\ {\bf #1},   #2 (#3)}
\def\mpl#1#2#3{Mod.\ Phys.\ Lett.\ {\bf A#1},  #2 (#3)}
\def\rep#1#2#3{Phys.\ Rept.\        {\bf #1},   #2 (#3)}
\def\sci#1#2#3{Science             {\bf #1},   #2 (#3)}
\def\astro#1#2#3{Astrophys.\ J.\   {\bf #1},   #2 (#3)}
\def\epj#1#2#3{Eur.\ Phys.\ J.\   {\bf C#1},   #2 (#3)}
\def\jhep#1#2#3{JHEP              {\bf #1},   #2 (#3)}
\def\ptp#1#2#3{Prog.\ Theor.\ Phys.\ {\bf #1}, #2 (#3)}
\def\ijmp#1#2#3{Int.\ J.\ Mod.\ Phys.\ {\bf A#1}, #2 (#3)}


\bibitem{minko}
P.~Minkowski,
Phys.\ Lett.\ B {\bf 67} 421 (1977); M. Gell-Mann, P. Ramond and
R. Slansky, in {\it  Supergravity}, eds.\ P. Van Nieuwenhuizen and
D. Freedman (North-Holland, Amsterdam, 1979),p.~315; T. Yanagida,
in {\it Proceedings of the Workshop on the Unified Theory
  and the Baryon Number in the Universe}, eds.\ O. Sawada and
  A. Sugamoto (KEK, Tsukuba, 1979), p.~95;
 S.L. Glashow, in {\it Quarks and Leptons}, eds.\ M. L\'evy et al.,
(Plenum, 1980, New-York), p. 707.
\bibitem{lrsm}
R.N.~Mohapatra and G. Senjanovi\'{c}, Phys.\ Rev.\ Lett.\ {\bf
44}, (1980) 912.

\bibitem{Zreview}
Some reviews are, J.L. Hewett and T.G. Rizzo,
\rep{183}{193}{1989}; P. Langacker, M.x. Luo and A.K. Mann,
\rmp{64}{87}{1992} A. Leike, \rep{317}{143}{1999}; K.S. Babu and
C. Kolda in S. Eidelman {\it et al} (PDG), \plb{592}{1}{2004}.

\bibitem{nonexotic}
For more general studies, see, T. Appelquist, B.A. Dobrescu and
A.R. Hopper, \prd{68}{035012}{2003}; M. Carena, A. Daleo, B.A.
Dobrescu and T.M.P. Tait, hep-ph/0408098; J. Kang, P. Langacker
and T. Li, hep-ph/0411404.


\bibitem{keung}
W.Y. Keung and G. Senjanovic, \prl{50}{1427}{1983}.

\bibitem{fukugita}
M. Fukugita and T. Yanagida, \plb{174}{45}{1986}.

\bibitem{ibarra}
S. Davidson and A. Ibarra, \plb{535}{25}{2002}; W. B\"uchm\"uller,
P. Di Bari, M. Pl\"umacher, \plb{547}{128}{2002};
\npb{665}{445}{2003}.

\bibitem{resonantL}
A. Pilaftsis, \prd{56}{5431}{1997}.


\bibitem{softL}
Y. Grossman, T. Kashti, Y. Nir and E. Roulet,
\prl{91}{251801}{2003}, hep-ph/0407063; G. D'Ambrosio, G.F.
Giudice and M. Raidal, \plb{575}{75}{2003}.

\bibitem{chun}
E.J. Chun, \prd{69}{117303}{2004}.

\bibitem{kitano}
 Y. Grossman, R. Kitano and H. Murayama, hep-ph/0504160.

\bibitem{tev}
A. Pilaftsis and T.E.J. Underwood, \npb{692}{303}{2004}; T.
Hambye, J. March-Russell and S.M. West, \jhep{0407}{070}{2004}; L.
Boubekeur, T. Hambye and G. Senjanovic, \prl{93}{111601}{2004}.

\bibitem{king}
N. Sahu and U.A. Yajnik, \prd{71}{023507}{2005}; S.F. King and T.
Yanagida, hep-ph/0411030.

\bibitem{apos}
A. Pilaftsis and T.E.J. Underwood, hep-ph/0506107.


\bibitem{btol}
V.A. Kuzmin, V.A. Rubakov and M.E. Shaposhnikov,
\plb{155}{36}{1985}; S.Y. Khlebnikov and M.E. Shaposhinikov,
\npb{308}{885}{1988}; P. Arnold and L.D. McLerran,
\prd{36}{581}{1987};  L. Carson, X. Li, L.D. McLerran and R.T.
Wang, \prd{42}{2127}{1990}.


\bibitem{futureZ}
M. Battaglia, S. De Curtis, D. Dominici, A. Ferrari and J.
Heikkinen, hep-ph/0101114;  S. Godfrey, hep-ph/0201093; M.
Dittmar, A-S. Nicollerat and A. Djouadi, hep-ph/0307020; A.
Ferrari, \prd{65}{093008}{2002}.


\bibitem{schneider} O. Schneider, review in S. Eidelman {\it et al}
(PDG), \plb{592}{1}{2004}.


\end{thebibliography}
\end{document}